\documentclass[aps,prl,twocolumn,superscriptaddress,footinbib,longbibliography]{revtex4-1} 

\usepackage{longtable}
\usepackage{morefloats}
\usepackage[dvips]{graphicx}
\usepackage{color}
\usepackage{epsfig,graphicx,amsfonts,amsbsy}
\usepackage{amsmath,amsfonts,amsthm,amssymb}
\usepackage{appendix}
\usepackage{makeidx}
\usepackage{url}
\usepackage{verbatim}
\usepackage[bookmarksnumbered,pdfpagelabels=true,plainpages=false,colorlinks=true,linkcolor=blue,citecolor=red,urlcolor=blue]{hyperref}
\usepackage[rightcaption]{sidecap}
\usepackage{array}
\usepackage{multirow}
\usepackage{tabularx}
\usepackage{braket} 
\usepackage{dsfont}
\usepackage{bbold}
\usepackage{etoolbox}
\usepackage{comment}

\AtBeginDocument{%
    \newwrite\bibnotes
    \def\bibnotesext{Notes.bib}
    \immediate\openout\bibnotes=\jobname\bibnotesext
    \immediate\write\bibnotes{@CONTROL{REVTEX41Control}}
    \immediate\write\bibnotes{@CONTROL{%
    apsrev41Control,author="08",editor="1",pages="0",title="0",year="1"}}
     \if@filesw
     \immediate\write\@auxout{\string\citation{apsrev41Control}}%
    \fi
}%

\newcommand{\bs}[1]{\boldsymbol{#1}}

\def\Br{{\bs{r}}}
\def\Bv{{\bs{v}}}

\def\Bs{{\bs{s}}}
\def\Bsigma{{\bs{\sigma}}}

\def\xhat{{ \bs{\hat{x}} }}
\def\yhat{{ \bs{\hat{y}} }}
\def\zhat{{ \bs{\hat{z}} }}

\def\Bn{{ \bs{n} }}

\def\CalE{{\mathcal{E}}}
\def\CalS{{\mathcal{S}}}
\def\CalD{{\mathcal{D}}}
\def\CalL{{\mathcal{L}}}
\def\CalJ{{\mathcal{J}}}
\def\CalK{{\mathcal{K}}}

\begin{document}

\title{Majorana Bound States Induced by Antiferromagnetic Skyrmion Textures}

\author{Sebasti{\'a}n A. D{\'i}az}
\affiliation{Department of Physics, University of Basel, Klingelbergstrasse 82, CH-4056 Basel, Switzerland}

\author{Jelena Klinovaja}
\affiliation{Department of Physics, University of Basel, Klingelbergstrasse 82, CH-4056 Basel, Switzerland}

\author{Daniel Loss}
\affiliation{Department of Physics, University of Basel, Klingelbergstrasse 82, CH-4056 Basel, Switzerland}

\author{Silas Hoffman}
\affiliation{Department of Physics, University of Florida, Gainesville, Florida 32611, USA}
\affiliation{Quantum Theory Project, University of Florida, Gainesville, Florida 32611, USA}
\affiliation{Center for Molecular Magnetic Quantum Materials, University of Florida, Gainesville, Florida 32611, USA}
\affiliation{Department of Physics, University of Basel, Klingelbergstrasse 82, CH-4056 Basel, Switzerland}

\date{\today}
	
\begin{abstract}
Majorana bound states are zero-energy states predicted to emerge in topological superconductors and intense efforts seeking a definitive proof of their observation are still ongoing. A standard route to realize them involves antagonistic orders: a superconductor in proximity to a ferromagnet. Here we show this issue can be resolved using antiferromagnetic rather than ferromagnetic order. We propose to use a chain of antiferromagnetic skyrmions, in an otherwise collinear antiferromagnet, coupled to a bulk conventional superconductor as a novel platform capable of supporting Majorana bound states that are robust against disorder. Crucially, the collinear antiferromagnetic region neither suppresses superconductivity nor induces topological superconductivity, thus allowing for Majorana bound states localized at the ends of the chain. Our model introduces a new class of systems where topological superconductivity can be induced by editing antiferromagnetic textures rather than locally tuning material parameters, opening avenues for the conclusive observation of Majorana bound states.
\end{abstract}

\maketitle

%%%%%%%%%%%%%%%%%%%%%%%%%%%%%%%%%%%%%%%%%%%%%%%%%%%%%%%%%%
%%%%%%%%%%%%%%%%%%%%%%%%%%%%%%%%%%%%%%%%%%%%%%%%%%%%%%%%%%

One-dimensional topological superconductors host zero-energy states localized at their ends called Majorana bound states (MBSs)~\cite{kitaevPU01,aliceaRPP12,pawlakPPNP19,Prada2020}. An important driving force in the research of MBSs stems from their non-Abelian exchange statistics~\cite{ivanovPRL01,nayakRMP08,aliceaNATP11}, a key property that makes them attractive for their potential use in topological quantum computing~\cite{Stern2013,Sarma2015}. Promising experimental signatures of zero-energy states consistent with MBSs have been reported in nanowires~\cite{mourikSCI12,dasNATP12,dengSCI16,Suominen2017,marcusNATP20} and atomic chains~\cite{nadj-pergeSCI14,rubyPRL15,pawlakNPJ16,Kim2018}, which are predicted to realize one-dimensional topological superconductivity~\cite{lutchynPRL10,oregPRL10,klinovajaPRL13,vazifehPRL13,brauneckerPRL13,pientkaPRB13}. However, due to the inherent static nature of these setups, they do not easily lend themselves to the ultimate test of measuring exchange statistics.

\begin{figure}[!t]
\includegraphics[width=\columnwidth]{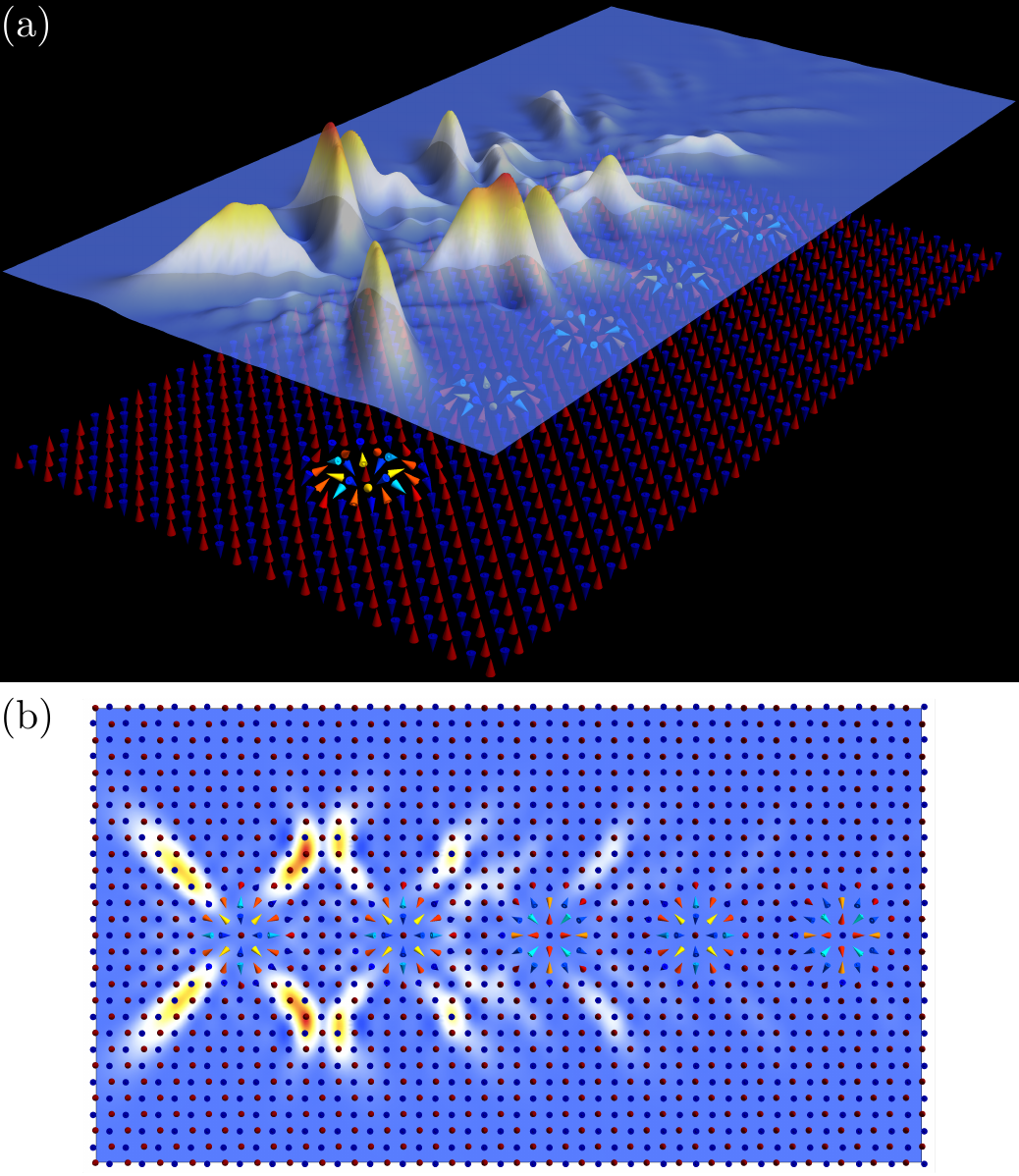}
\caption{Antiferromagnetic skyrmion chains induce Majorana bound states.
(a)~Probability density of a Majorana bound state (top) localized at one end of a chain of antiferromagnetic skyrmions embedded in a collinear antiferromagnet (bottom).
(b)~Top view of the magnetic texture and Majorana bound state probability density. The partner Majorana bound state is localized at the other end of the chain (not shown). Texture parameters are detailed in Methods.
}
\label{tex}
\end{figure}

An alternative route toward topological superconductivity is to couple a bulk superconductor to a noncollinear magnetic texture~\cite{brauneckerPRB10,klinovajaPRL12,kjaergaardPRB12,Egger2012,klinovajaPRL13,KlinovajaPRX2013,vazifehPRL13,brauneckerPRL13,poyhonenPRB14,Matos-Abiague2017,Desjardins2019,steffensenCM20,rexCM20}, such as stable topological defects known as skyrmions~\cite{muhlbauerSCI09,yuNAT10,Poyhonen2016,Gungordu2018,Rex2019,Garnier2019,kubetzkaPRM20}. Recently, single ferromagnetic (FM) skyrmions with large topological charge~\cite{guangPRB16} and FM skyrmion lattices~\cite{nagaosaPRB13,mascotCM20,mohantaCM20} coupled to a superconductor have been predicted to support MBSs and Majorana edge modes, respectively. Because of their topological stability, skyrmions are themselves particle-like objects which can be moved using temperature gradients~\cite{Mochizuki2014}, electric currents~\cite{Jonietz1648,Jiang283,Jiang_2016,Woo_2016}, or magnetic field gradients from magnetic-force-microscope tips~\cite{Casiraghi_2019}. Their mobility opens the door to the assembly and dynamical manipulation of structures that can facilitate the efforts to measure the exchange statistics of MBSs. A major hindrance, though, is that the ability to move the skyrmions necessitates embedding them into a two-dimensional collinear FM background, which destroys the proximity-induced superconducting gap, delocalizing the zero-energy states.

\clearpage
To resolve this issue, here we consider a collinear antiferromagnetic (AFM) background into which a chain of AFM skyrmions~\cite{Barker2016,Zhang2016} is embedded, see Fig.~\ref{tex}. This is an ideal system because AFM skyrmions can be stabilized without external magnetic fields and the collinear AFM order does not have a harmful effect on superconductivity. We discover that in precisely the AFM system, we can modulate between topological and nontopological regions by editing between noncollinear and collinear AFM order, respectively. We also show that its richer phase diagram exhibits more possibilities to tune into the topological superconducting phase. Moreover, under equivalent conditions, we find that the minimum length necessary to ensure well-localized MBSs is smaller in antiferromagnetic skyrmion chains (ASCs) than in one-dimensional AFM spin helices.\\

%%%%%%%%%%%%%%%%%%%%%%%%%%%%%%%%%%%%%%%%%%%%%%%%%%%%%%%%%%
%%%%%%%%%%%%%%%%%%%%%%%%%%%%%%%%%%%%%%%%%%%%%%%%%%%%%%%%%%

\noindent{\bf \large Results}

\noindent{\bf Model.}
Using a lattice model, we describe our system with the Hamiltonian,
\begin{align}\nonumber
&H = - \sum_{\Br , \sigma} \mu \, c_{\Br \sigma}^\dag c_{\Br \sigma} - \sum_{\langle\Br,\Br'\rangle , \sigma} t \, c_{\Br \sigma}^\dag c_{\Br' \sigma} \\
&+ \sum_{\Br} \Delta (c_{\Br \uparrow}^\dag c_{\Br \downarrow}^\dag + c_{\Br \downarrow} c_{\Br \uparrow}) + \sum_{\Br,\mu,\nu} J c_{\Br \mu}^\dag \Bn_{\Br} \cdot \Bsigma_{\mu \nu} \, c_{\Br \nu} \,,
\label{ham}
\end{align}
where $c_{\Br\sigma}^\dagger$ creates an electron at lattice site $\Br$ with spin $\sigma$ in a thin-film magnetic conductor that is coupled to a superconductor, which induces a gap, $\Delta$, by proximity effect. The electronic properties of this conductor are governed by the chemical potential, $\mu$, and nearest neighbor hopping amplitude, $t$, while the direction of the magnetic texture, $\Bn_{\Br}$, is coupled via the exchange interaction, with strength $J$, to the local spin of the conductor, $\Bs_{\Br} = (1/2) \sum_{\mu,\nu} c_{\Br \mu}^\dag \Bsigma_{\mu \nu} \, c_{\Br \nu}$, where $\Bsigma$ is the vector of Pauli matrices. For simplicity, the lattice constant is henceforth taken as unity.\\

%%%%%%%%%%%%%%%%%%%%%%%%%%%%%%%%%%%%%%%%%%%%%%%%%%%%%%%%%%
%%%%%%%%%%%%%%%%%%%%%%%%%%%%%%%%%%%%%%%%%%%%%%%%%%%%%%%%%%

\noindent{\bf Advantages of an antiferromagnet.}
Although a natural point of departure for topological superconductivity may start with materials supporting a FM exchange interaction and, likewise, FM skyrmion chains, the critical reader could be weary of such a system for several reasons. First, in order to stabilize FM skyrmions, a magnetic field is conventionally applied perpendicular to the surface and could destroy the superconductivity. Second, if the texture manages to survive the conditions to support skyrmions, a collinear FM background will surely destroy the superconducting correlations at least for a modest value of the exchange interaction as compared to the AFM case. Lastly, if we are able to generate such a chain, because FM skyrmions naturally repel, maintaining such a chain would require pinning of the individual skyrmions with a local magnetic field or impurity to locally enhance the anisotropy.

Because in the following analysis we are interested in localized states supported by skyrmions in a collinear magnetic background, it is imperative that the spectrum remains gapped.  However, in an infinite collinear FM layer coupled to a superconductor, Eq.~(\ref{ham}) with $\Bn_\Br$ constant, the gap vanishes when $J\geq\Delta$. Consequently, any would-be localized states at the ends of a ferromagnetic skyrmion chain are delocalized throughout the superconductor under the collinear ferromagnetic region. In contrast, the spectral gap of a superconductor coupled to a collinear antiferromagnet closes only when $J=\sqrt{\Delta^2+\mu^2}$. Thus, although it is hopeless to realize MBSs generated by ferromagnetic skyrmions residing in a larger collinear background, as we show below, chains of antiferromagnetic skyrmions in a antiferromagnetic collinear background can host MBSs at the their ends.\\

%%%%%%%%%%%%%%%%%%%%%%%%%%%%%%%%%%%%%%%%%%%%%%%%%%%%%%%%%%
%%%%%%%%%%%%%%%%%%%%%%%%%%%%%%%%%%%%%%%%%%%%%%%%%%%%%%%%%%

\noindent{\bf Antiferromagnetic skyrmion textures.} 
For a \textit{realistic} treatment of the skyrmionic textures, we consider the free energy of the localized spins whose classical minimum configuration determines $\Bn_{\Br}$. Including AFM exchange and interfacial Dzyaloshinskii-Moriya interactions, as well as easy-axis anisotropy, we use atomistic spin simulations to numerically obtain an ASC (Fig.~\ref{tex}) as a metastable spin configuration. We emphasize here the complete absence of any external magnetic field (which, for the ferromagnetic case, would be needed to stabilize a skyrmion texture).

Below we also employ \textit{artificial} skyrmionic textures constructed using the following simple model. A single AFM skyrmion on the square lattice centered at the origin is parametrized by $\Bn_{\Br}=(-1)^{i+j}[\sin(k_S r) \cos\varphi(\Br), \sin(k_S r) \sin\varphi(\Br), \cos(k_S r)]^T$ for $r\leq R_S$ and $\Bn_{\Br}=[0,0,(-1)^{i+j}]^T$ otherwise, where $\Br = i \xhat + j \yhat$ and $\tan \varphi(\Br) = j/i$ with $i$ and $j$ being integers.  Here $k_S=\pi/(R_S-1)$ and $R_S$ is the skyrmion radius. This model can be immediately generalized to a chain of AFM skyrmions with a spacing $\CalS$ between edges of two adjacent skyrmions. By tuning $R_S$ and $\CalS$, \textit{artificial} skyrmionic textures can be brought to remarkably good agreement with \textit{realistic} textures obtained by atomistic spin simulations.\\ 

%%%%%%%%%%%%%%%%%%%%%%%%%%%%%%%%%%%%%%%%%%%%%%%%%%%%%%%%%%
%%%%%%%%%%%%%%%%%%%%%%%%%%%%%%%%%%%%%%%%%%%%%%%%%%%%%%%%%%

\begin{figure}[t]
\includegraphics[width=\columnwidth]{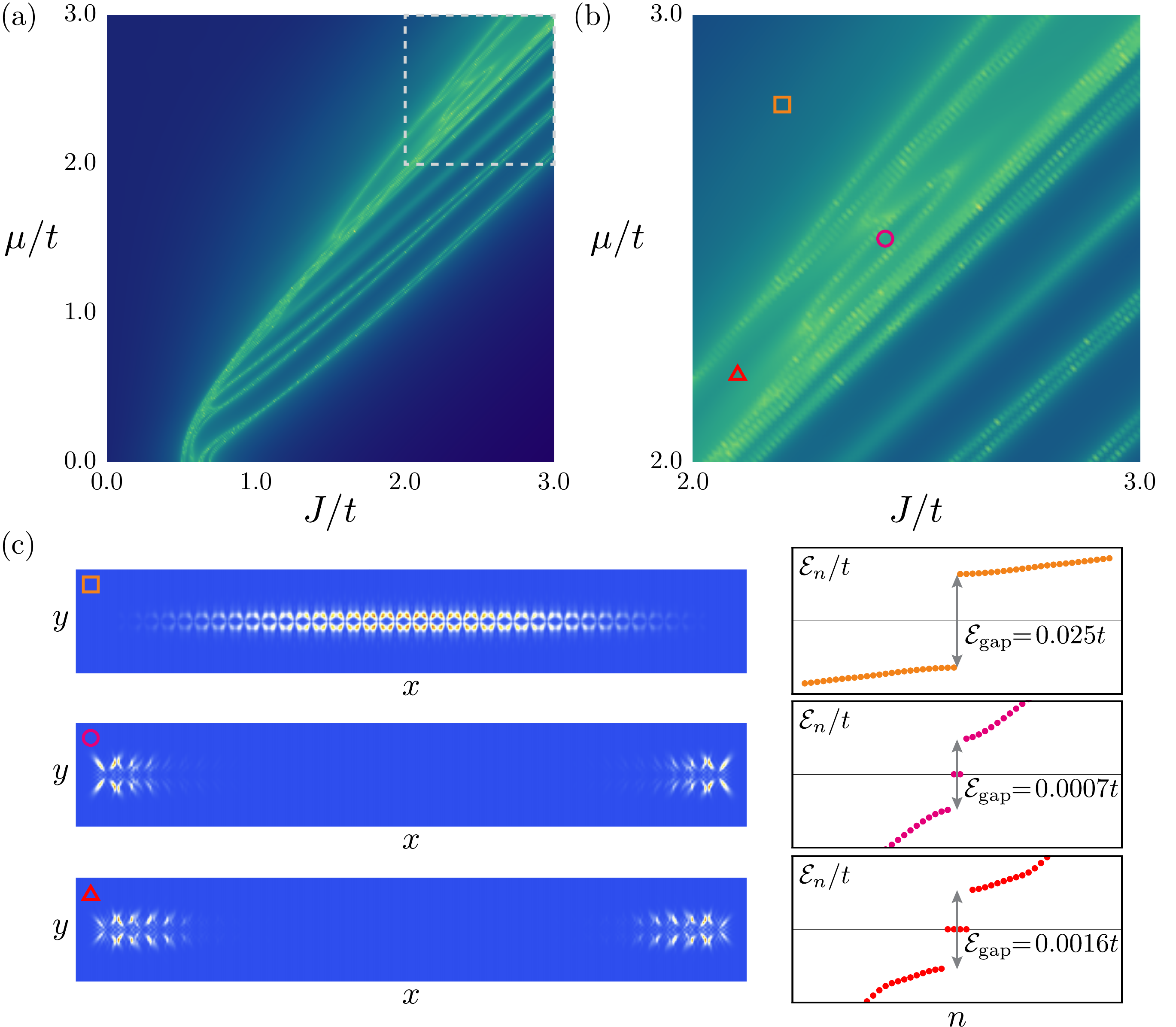}
\caption{Topological phases induced by an antiferromagnetic skyrmion chain.
(a)~Topological phase diagram as a function of the chemical potential $\mu$ and exchange coefficient $J$ for $\Delta=0.5 t$. The logarithmically-scaled color code encodes the energy gap, $\CalE_\textrm{gap}$, and the green curves denote $\CalE_\textrm{gap} \approx 0$.
(b)~Blowup of the region enclosed by the dashed gray square in (a). Highlighted are selected phases supporting no MBSs (orange square, $\{\mu,J\}=\{2.8t,2.2t\}$), two MBSs (magenta circle, $\{\mu,J\}=\{2.5t,2.43t\}$), and four MBSs (red triangle, $\{\mu,J\}=\{2.2t,2.1t\}$).
(c)~Probability density of the lowest nonnegative energy state (left) and energy spectrum (right) of the selected phases indicated in (b) for a chain composed of 37 antiferromagnetic skyrmions. The texture parameters are the same as in Fig.~\ref{tex}.
}
\label{pd}
\end{figure}

\noindent{\bf Induced Majorana bound states.}
To confirm that the ASC is able to induce MBSs, we construct a phase diagram with control parameters $\mu$ and $J$, choosing $\Delta=0.5 t$ for numerical simplicity, by finding the bulk gap closing points in the periodic realistic ASC. In Fig.~\ref{pd}(a) we plot the energy gap, $\CalE_\textrm{gap}$, scaled logarithmically where green color indicates $\CalE_\textrm{gap} \approx 0$. There is a region in the phase diagram, enclosed by the dashed gray square, in which changes in the gap size are dense. A blowup of this region is shown in Fig.~\ref{pd}(b) where three points have been highlighted. To characterize the corresponding phases, in Fig.~\ref{pd}(c) we plot the probability density of their lowest nonnegative energy state, $|\Psi_{\CalE_0}(\Br)|^2$, and their energy spectrum found for a finite-size chain of AFM skyrmions. Indeed, we find topologically distinct phases [separated by (green) gap closing lines], in which the ASC induces zero, two, and four MBSs, respectively. The lobular structure exhibited by the probability density of the MBSs is reminiscent of that of the lowest-energy state induced by single isolated AFM skyrmions (see Supplementary Note 1). It is also similar to the clover shape of Yu-Shiba-Rusinov states of magnetic adatoms on a superconducting surface measured by scanning tunneling spectroscopy~\cite{Ruby2018}. \\

%%%%%%%%%%%%%%%%%%%%%%%%%%%%%%%%%%%%%%%%%%%%%%%%%%%%%%%%%%
%%%%%%%%%%%%%%%%%%%%%%%%%%%%%%%%%%%%%%%%%%%%%%%%%%%%%%%%%%

\noindent{\bf Stability of Majorana bound states.}
One of the key properties of MBSs is that they remain at zero energy in the presence of disorder. In the system studied here, impurities in the sample could give rise to local variations in the electronic properties of a material, i.e. parameters of our model, such as the chemical potential and the exchange interaction. Another possible source of disorder is the underlying skyrmionic magnetic texture. For instance, randomly positioned impurities could locally deform the magnetic texture. Furthermore, skyrmionic magnetic textures at high temperatures (room temperature) can be regarded as an ensemble average over small random deformations throughout the entirety of the pristine zero-temperature texture.

We have studied the effect of disorder in the electronic properties as well as the texture on the MBSs localized at the ends of AFM skyrmion chains. To determine the effect of each possible source of disorder independently of the others, we considered the following four disorder models: randomly distributed chemical potential $\mu$, randomly distributed exchange interaction $J$, random flips of the magnetic moments of the texture, and small random deviations of the moments across the whole texture. The details of these disorder studies can be found in Supplementary Note 2. Ultimately,  the MBSs are robust to disorder in the electronic couplings $\mu$ and $J$, similar to MBSs arising in quantum wires~\cite{brouwerPRB11}, and are more sensitive to deformations in the magnetic texture.\\

%%%%%%%%%%%%%%%%%%%%%%%%%%%%%%%%%%%%%%%%%%%%%%%%%%%%%%%%%%
%%%%%%%%%%%%%%%%%%%%%%%%%%%%%%%%%%%%%%%%%%%%%%%%%%%%%%%%%%

\noindent{\bf Effective Hamiltonian.}
A single AFM skyrmion residing in a collinear background hosts several localized low-energy states  (see Supplementary Note 1). When skyrmions are sufficiently close together, the localized states hybridize and form a band. We can approximate and study the lowest energy band by projecting Eq.~(\ref{ham}) onto the states closest to the chemical potential. That is, let $|\Psi^+_L\rangle$ ($|\Psi^-_L\rangle$) be the lowest positive (negative) energy state of a single skyrmion centered at $\Br=0$ and $|\Psi^+_R\rangle$ ($|\Psi^-_R \rangle$) be the analogous lowest positive (negative) energy state of a single skyrmion centered at $\Br=\CalL\xhat$ with $\CalL$ greater than the skyrmion radius. Upon projecting our original Hamiltonian onto these states, we obtain an effective Kitaev chain~\cite{kitaevPU01} with chemical potential $\mu_\textrm{eff}=\langle \Psi^+_L|H|\Psi^+_L\rangle$, nearest-neighbor hopping $t_\textrm{eff}=\langle \Psi^+_R|H|\Psi^+_L\rangle$, and superconducting pairing $\Delta_\textrm{eff}=\langle \Psi^-_R|H|\Psi^+_L\rangle$. Because we are only using the lowest energy quasiparticle and its hole partner and neglecting the effect of states hosted by next-to-nearest neighbor skyrmions (and beyond), this projection gives a rough diagnosis of the presence of a pair of MBSs, i.e. if $2|t_ \textrm{eff}|>|\mu_\textrm{eff}|$, which one can confirm or disprove by explicitly calculating the spectrum and wave functions of the corresponding open skyrmion chain.\\

%%%%%%%%%%%%%%%%%%%%%%%%%%%%%%%%%%%%%%%%%%%%%%%%%%%%%%%%%%
%%%%%%%%%%%%%%%%%%%%%%%%%%%%%%%%%%%%%%%%%%%%%%%%%%%%%%%%%%

\begin{figure}[t]
\includegraphics[width=\columnwidth]{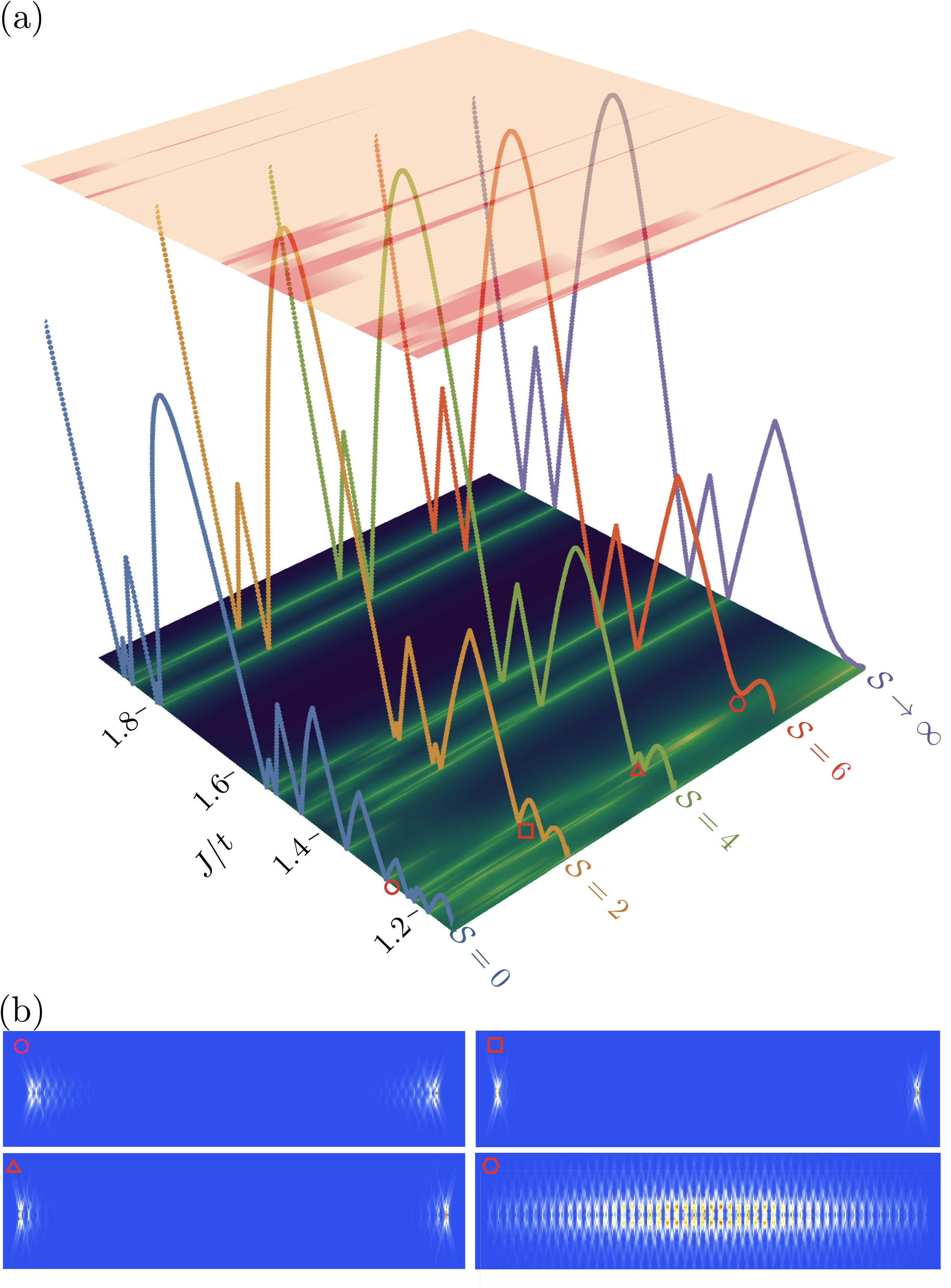}
\caption{Effect of spacing between adjacent skyrmions.
(a)~Topological phase diagram of a periodic chain of antiferromagnetic skyrmions, of radius $R_S=4$ with $\mu=1.0 t$ and $\Delta=0.5 t$, with control parameters $J$ and the spacing between adjacent skyrmions $\CalS$. The curves denote the energy gap at the specified values of $\CalS$ as a function of $J$. The bottom face of the cube interpolates the logarithmically-scaled gap values of the curves along the $\CalS$ axis. The top face of the cube indicates the topological number of the effective Kitaev chain, described in the text, in which red (off-white) corresponds to one (zero).
(b)~Probability density of the lowest nonnegative energy state at selected points marked in the phase diagram (circle: $\CalS=0$; square: $\CalS=2$; triangle: $\CalS=4$; and hexagon: $\CalS=6$) for an open chain of 50 antiferromagnetic skyrmions. MBSs remain robust against a moderate spacing increase.
}
\label{sp1}
\end{figure}

\noindent{\bf Skyrmion density.}
Although distances between skyrmions are fixed in the realistic texture by the spin simulation parameters, we can study how the density of skyrmions in the artificial texture affects the formation of MBSs by increasing the spacing between adjacent skyrmions, $\CalS = \CalL - 2 (R_S - 1)$. As a concrete case study, in Fig.~\ref{sp1}(a) we plot the gap size for a periodic chain of skyrmions as a function of $J$ in a series of curves. The frontmost curve corresponds to a chain with $\CalS=0$ and successive curves, moving back, increase the spacing by two until the penultimate curve; the last curve corresponds to the limit $\CalS \to \infty$ of an isolated, single skyrmion. The bottom face of the box is a phase diagram as a function of $J$ interpolated along the $\CalS$ axis. The top face of the box indicates if the corresponding point of the phase diagram below is expected to be topological (red) or nontopological (off-white) as predicted by the effective Hamiltonian. The densest skyrmion chain, $\CalS=0$, supports MBSs near $J\in\{1.55t, 1.825t,1.9t\}$, and in the range $J\in[1.37t,1.47t]$. Increasing $\CalS$ noticeably decreases the range of $J$ supporting a topological phase which asymptotes to the case of an isolated skyrmion where the gap closes with the system remaining nontopological.  The range $J\in[1.24t,1.29t]$ is also topological for $\CalS=0$ with a decreasing range of $J$ for increasing $\CalS$ but, in contrast, the gap closing is lifted for sufficiently large $\CalS$, e.g. $\CalS=6$ near $J=1.21t$, wherein the topological phase changes into the nontopological one. We conclude that, although a skyrmion sufficiently displaced from its neighbors in the chain will break a single topological chain into two topological pieces, MBSs at the chain ends can still enjoy protection against moderate skyrmion displacements. Furthermore, we predict that the denser the packing of skyrmions along the chain, the denser the topological phase in parameter space.\\

%%%%%%%%%%%%%%%%%%%%%%%%%%%%%%%%%%%%%%%%%%%%%%%%%%%%%%%%%%
%%%%%%%%%%%%%%%%%%%%%%%%%%%%%%%%%%%%%%%%%%%%%%%%%%%%%%%%%%

\noindent{\bf Experimental realization and observation.}
Recent experiments have demonstrated the presence of FM skyrmions in Fe/Ir thin films grown on a Re substrate, the latter supporting superconductivity at sufficiently low temperatures~\cite{kubetzkaPRM20}, though, the coexistence of superconductivity with noncollinear ferromagnetism has not yet been established. Simultaneously, evidence consistent with topological superconductivity has been measured in nanowires overgrown with interlaced FM EuS and superconducting Al~\cite{marcusNATP20}. These systems are evidence of the experimental expertise necessary to realize magnetic-superconducting heterostructures which could guide the engineering of our proposed system.

One route to experimentally realizing our setup is a heterostructure of layered transition metal dichalcogenides (TMDs). TMDs enjoy a large spin-orbit coupling, are amenable to being stacked, and their chemical potential can be shifted by the application of a back gate. Recent advances in the synthesis of TMDs have uncovered materials that can host superconducting or AFM order, though not concurrently.  In particular, NiSe$_2$ is a promising candidate for the superconducting layer as it sustains superconductivity even as monolayer~\cite{xiNATP16}. On the other hand, Fe$_{1/3}$NiS$_2$ exhibits AFM~\cite{nairNATM20} order though noncollinear AFM order remains elusive. Analogous to conventional FM materials~\cite{Fert1980,Moreau-Luchaire2016}, we propose interfacing the antiferromagnet with a layer of heavy metals, e.g. Ir, Pa, or Pt, which could enhance the spin-orbit interaction and ultimately provide a strong Dzyaloshinskii-Moriya interaction to stabilize skyrmions. A trilayer of Fe$_{1/3}$NiS$_2|$Ir$|$NiSe$_2$ would allow injection or deletion of AFM skyrmions in the magnetic layer, stabilized by the heavy metal, which induces an effective magnetic exchange interaction within the NiSe$_2$ superconducting layer. Because the latter can be back gated~\cite{dvitPRL19}, the topological phase space of the heterostructure, which we predict to be dense, can be scanned or tuned to the topologically nontrivial regime.

\begin{figure}[t]
\includegraphics[width=\columnwidth]{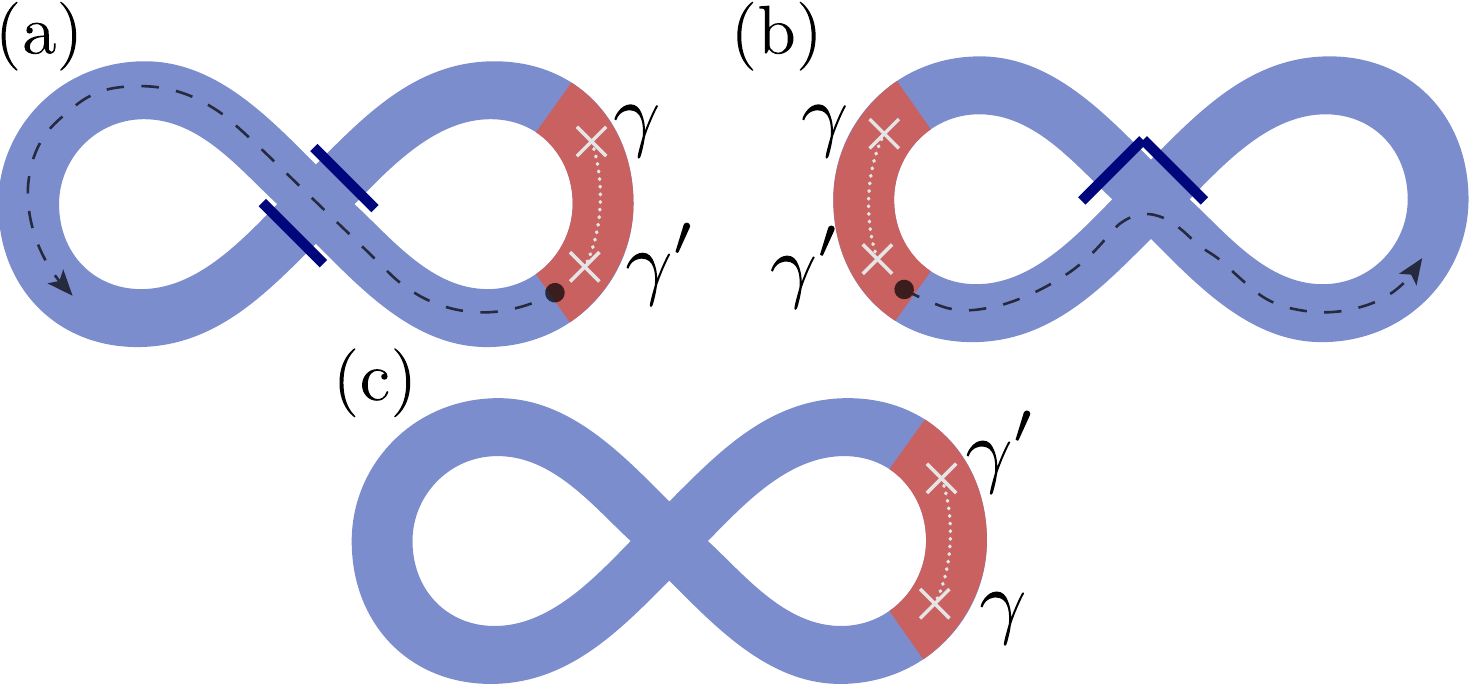}
\caption{Exchanging MBSs on a topological racetrack.
(a)~MBSs, $\gamma$ and $\gamma'$ residing at the ends of an antiferromagnetic skyrmion chain (red) can be shuffled to the other side of the structure, through the collinear antiferromagnetic background (blue), utilizing a path through the cross.
(b)~The same procedure can move the MBSs back to the right via the lower leg of the structure.
(c)~The result is an exchange of $\gamma$ and $\gamma'$.
}
\label{braid}
\end{figure}

Two-dimensional systems hosting more complicated geometries of chains could provide a path for the identification of MBSs. Rather than a simple straight chain of skyrmions, consider a quasi-one-dimensional figure-eight track which supports antiferromagnetism and conventional superconductivity described by Eq.~(\ref{ham}). A chain of skyrmions located initially at the right curve in the track supports two MBSs, $\gamma$ and $\gamma'$ at each end [Fig.~\ref{braid}(a)]. This chain can be pinned, e.g. by a magnetic tip or a local impurity. Then the chain can be moved, by a spin current for instance, to the other side by going through the vertex of the quasi-one-dimensional structure [Fig.~\ref{braid}(b)]. Shuffling the skyrmions back to the right using only the lower leg of the structure consequently exchanges the MBSs [Fig.~\ref{braid}(c)] and imprints an overall phase which can be measured by an ancillary state. This can be generalized to more skyrmion chains and quasi-one-dimensional structures with additional handles. We can effect braiding of any two MBSs by analogous shuffling in what may be deemed a kind of `topological racetrack memory.'\\

%%%%%%%%%%%%%%%%%%%%%%%%%%%%%%%%%%%%%%%%%%%%%%%%%%%%%%%%%%
%%%%%%%%%%%%%%%%%%%%%%%%%%%%%%%%%%%%%%%%%%%%%%%%%%%%%%%%%%

\noindent{\bf \large Discussion}\\
So far, AFM skyrmions have not been observed in conventional AFM materials. However, real-space detection of skyrmions has been recently reported in ferrimagnets~\cite{Woo2018} and synthetic antiferromagnets~\cite{Legrand2019,Dohi2019,Li2020,Wang2020}. The experimental assembly of chains of, albeit FM, skyrmions has already been achieved~\cite{Du2015,Hou2018,Casiraghi_2019}.

Our model, Eq.~(\ref{ham}), can also describe the experimentally different setup of a lattice of magnetic atoms residing on the surface of a superconductor. In two dimensions such systems are predicted to support Majorana edge modes as in the case of so-called Shiba lattices~\cite{Li2016} and as recently observed in van der Waals heterostructures~\cite{Kezilebieke2020}. One-dimensional chains of magnetic atoms on conventional superconductors have also been experimentally realized and zero-energy states localized at the ends of the chains have been observed~\cite{nadj-pergeSCI14,pawlakNPJ16,Kim2018}. Owing to interfacial Dzyaloshinskii-Moriya interactions, the atomic chain in Ref.~\cite{Kim2018} orders in a FM spin helix, crucial for the appearance of MBSs. Moreover, a one-dimensional AFM spin helix can also support MBSs \cite{heimesPRB14}, as we explain analytically in Supplementary Note 3.

Even though the one-dimensional texture along the center of the ASC is an AFM spin helix, the former has an important advantage over the latter. Localized MBSs are guaranteed as long as their wave functions do not overlap, which could lead to their hybridization. For the same induced gap, which determines MBSs localization length, the minimum length necessary to ensure their localization is smaller in ASCs than in one-dimensional AFM spin helices, as we show in Supplementary Note 4. Intuitively, the lateral extension of ASCs provides additional spatial support for the MBSs wave function to spread out and shorten their localization length along the chain axis~\cite{zyuzinPRL13,Peng2015}.

Although our analysis has been done for chains composed of skyrmions whose topological number is strictly one, AFM skyrmions provide a route to connect our system to other skyrmion systems which are known to host MBSs, e.g. skyrmions with large topological charge~\cite{guangPRB16} and FM skyrmion lattices~\cite{nagaosaPRB13,mascotCM20}, by adiabatically deforming the magnetic texture. For instance, a single chain of skyrmions can be adiabatically deformed into a single skyrmion with large topological number. Likewise, multiple initially uncoupled chains of skyrmions can be slowly brought into proximity with each other to form a lattice. In both cases, it would be interesting to see how the spectrum and spatial distribution of the MBSs wave function evolve.

%%%%%%%%%%%%%%%%%%%%%%%%%%%%%%%%%%%%%%%%%%%%%%%%%%%%%%%%%%
%%%%%%%%%%%%%%%%%%%%%%%%%%%%%%%%%%%%%%%%%%%%%%%%%%%%%%%%%%

%\textit{Conclusions.---} 

In contrast to ferromagnets, antiferromagnets are capable of changing the local topological phase of an antiferromagnet$|$superconductor bilayer by deforming between collinear and noncollinear magnetic textures. In particular, chains of AFM skyrmions are capable of hosting MBSs at their ends, generating a rich phase diagram that depends on the material parameters and geometric details of the skyrmions, further increasing the degree of tunability into the topological superconducting phase. These end states are robust to fluctuations in the chemical potential and magnitude of the exchange interaction as well as small deformations in the magnetic texture. Our system offers a new platform in which local topological superconducting regions can be moved and modified by $\textit{in situ}$ manipulation of the magnetic order, and provides a potential route to measuring the exchange statistics and perform braiding of MBSs.\\

%%%%%%%%%%%%%%%%%%%%%%%%%%%%%%%%%%%%%%%%%%%%%%%%%%%%%%%%%%
%%%%%%%%%%%%%%%%%%%%%%%%%%%%%%%%%%%%%%%%%%%%%%%%%%%%%%%%%%

\noindent{\bf \large Methods} 

\noindent{\bf Realistic antiferromagnetic skyrmion textures.}
We use atomistic spin simulations to determine classical, metastable magnetic textures governed by the energy function
\begin{align}\label{eq:SpinH}\nonumber
\CalE_M =& \sum_{\Br} 
\CalJ \, \Bn_\Br \cdot \big( \Bn_{\Br + \xhat} + \Bn_{\Br + \yhat} \big) - \CalK (\Bn_\Br \cdot \zhat)^2 \\
& + \CalD ( \xhat \cdot \Bn_\Br \times \Bn_{\Br + \yhat} - \yhat \cdot \Bn_\Br \times \Bn_{\Br + \xhat} ) \,,
\end{align}
with $\Bn_\Br$ the direction of the classical spin at site $\Br$ on a square lattice, with the lattice constant taken as unity, located on the $xy$-plane, which includes nearest neighbor antiferromagnetic exchange $\CalJ$, interfacial Dzyaloshinskii-Moriya interaction $\CalD$, and easy-axis anisotropy $\CalK$. The phase diagram of single antiferromagnetic skyrmions modeled by a similar energy function has been recently discussed in Ref.~\cite{Bessarab2019}. An initial ``seed texture'' consisting of a chain of circular regions with downward pointing spins embedded in a collinear antiferromagnetic texture is relaxed employing the atomistic Landau-Lifshitz-Gilbert equation. The parameters that generate the texture used in Figs.~\ref{tex}-\ref{pd} are $\{\CalD/\CalJ,\CalK/\CalJ\} = \{0.475,0.4\}$. For a system size of $360 \times 45$ spins with free boundary conditions we obtained a chain of 37 antiferromagnetic skyrmions.\\

%%%%%%%%%%%%%%%%%%%%%%%%%%%%%%%%%%%%%%%%%%%%%%%%%%%%%%%%%%
%%%%%%%%%%%%%%%%%%%%%%%%%%%%%%%%%%%%%%%%%%%%%%%%%%%%%%%%%%

\noindent{\bf \large Acknowledgements}\\
We thank Ferdinand Schulz, Kirill Plekhanov, and Flavio Ronetti for useful discussions. This work was supported by the Swiss National Science Foundation and NCCR QSIT. This project received funding from the European Union's Horizon 2020 research and innovation program (ERC Starting Grant, Grant Agreement No. 757725). SH was also supported by the Center for Molecular Magnetic Quantum Materials, an Energy Frontier Research Center funded by the U.S. Department of Energy, Office of Science, Basic Energy Sciences under Award No. DE-SC0019330. 

%%%%%%%%%%%%%%%%%%%%%%%%%%%%%%%%%%%%%%%%%%%%%%%%%%%%%%%%%%
%%%%%%%%%%%%%%%%%%%%%%%%%%%%%%%%%%%%%%%%%%%%%%%%%%%%%%%%%%

%Replace
%\providecommand \doibase [0]{http://dx.doi.org/}%
%with
%\renewcommand{\doibase}[1]{https://dx.doi.org/\ifdefempty{#1}{}{#1}}%
%in the .bbl file to fix bug.

%\bibliography{mybib}

%

%%%%%%%%%%%%%%%%%%%%%%%%%%%%%%%%%%%%%%%%%%%%%%%%%%%%%%%%%%
%%%%%%%%%%%%%%%%%%%%%%%%%%%%%%%%%%%%%%%%%%%%%%%%%%%%%%%%%%

\setcounter{figure}{0}
\renewcommand{\thefigure}{ \arabic{figure}}
\renewcommand{\theHfigure}{\thefigure}
\renewcommand{\figurename}{Supplementary Figure}

\setcounter{equation}{0}
\renewcommand{\theequation}{S\arabic{equation}}
\renewcommand{\theHequation}{\theequation}

\clearpage

%%%%%%%%%%%%%%%%%%%%%%%%%%%%%%%%%%%%%%%%%%%%%%%%%%%%%%%%%%
%%%%%%%%%%%%%%%%%%%%%%%%%%%%%%%%%%%%%%%%%%%%%%%%%%%%%%%%%%

\section{Supplementary Note 1: States induced by a single antiferromagnetic skyrmion}

The probability density of the MBSs presented in the main text consists of an array of lobular patterns that fade toward the center of the antiferromagnetic skyrmion chain (ASC). We can understand how this peculiar spatial distribution emerges by looking at the electronic states supported by a single, isolated antiferromagnetic skyrmion. As we show in Supplementary Figure~\ref{LocalizedStates}, the probability density of the lowest positive energy state induced by the skyrmion exhibits a four-lobe structure. Comparison of Supplementary Figure~\ref{LocalizedStates} with the associated MBS wave functions in the main text [Fig.~\ref{pd}(c)] suggests that when the antiferromagnetic skyrmions are brought close to each other to assemble the ASC, the supported MBSs inherit their wave function from the lowest energy states and their hole partners. 

\begin{figure}[b!]
\centering
\includegraphics[width=0.85\columnwidth]{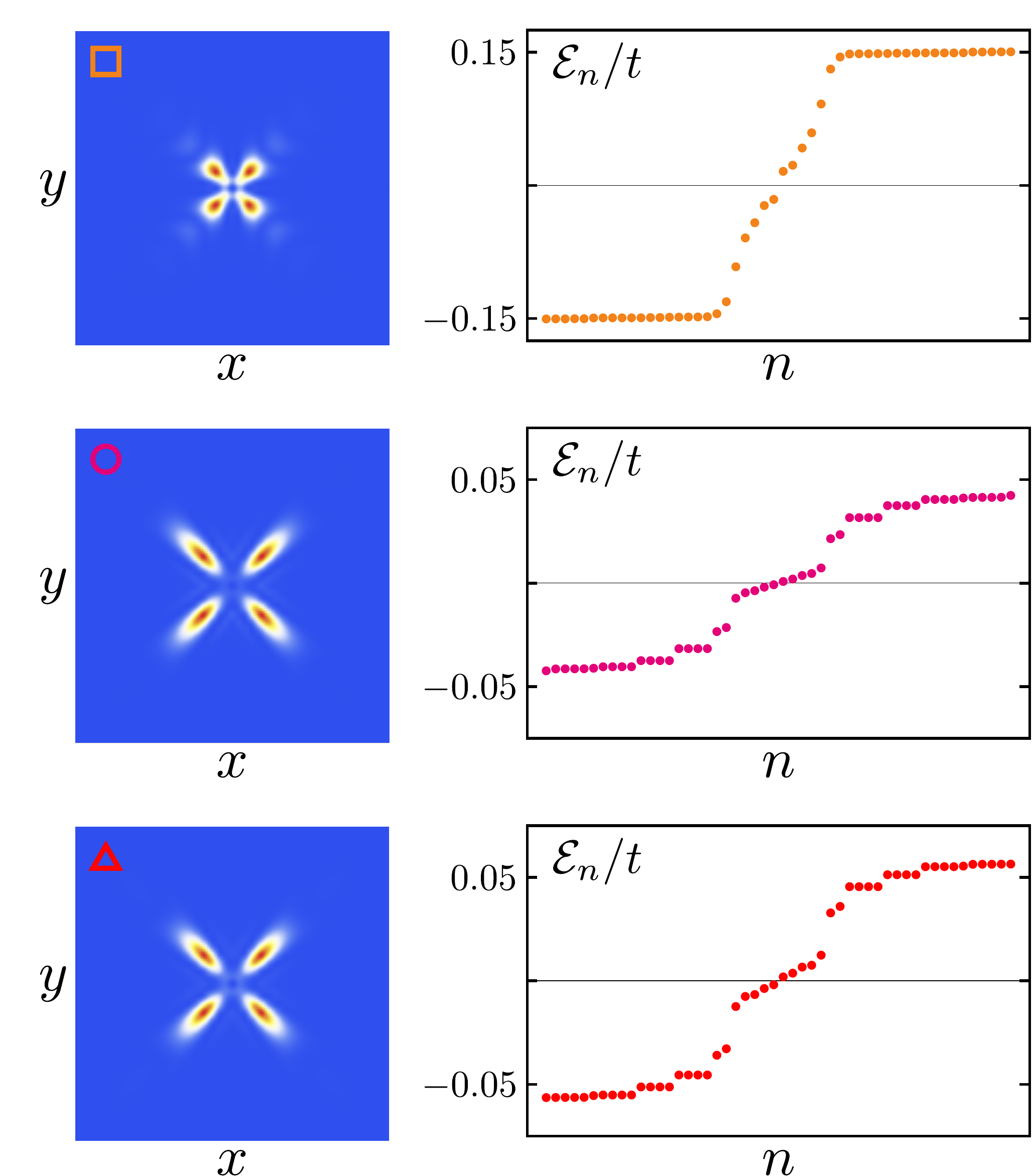}
\caption{Electronic states induced by a single antiferromagnetic skyrmion. Left panels: probability density of the lowest positive energy state for the same electronic parameters as in Fig.~\ref{pd} in the main text, indicated by the upper left symbol (orange square, magenta circle, and red triangle). Right panels: the corresponding energy spectra. The antiferromagnetic skyrmion is located at the center of a 45$\times$45 spin lattice generated using atomistic spin simulations and the magnetic parameters described in Methods. For all three cases, the lowest positive energy state shows a characteristic lobular structure centered at the antiferromagnetic skyrmion.}
\label{LocalizedStates}
\end{figure}

%%%%%%%%%%%%%%%%%%%%%%%%%%%%%%%%%%%%%%%%%%%%%%%%%%%%%%%%%%
%%%%%%%%%%%%%%%%%%%%%%%%%%%%%%%%%%%%%%%%%%%%%%%%%%%%%%%%%%

\section{Supplementary Note 2: Disorder analysis}

In the absence of disorder, a linear chain of 37 AFM skyrmions with electronic model parameters $\{ \mu, \Delta, J \} = \{ 2.0t, 0.5t, 1.92t \}$ supports four MBSs. First we studied the effect of disorder arising from spatial inhomogeneities in the chemical potential and the exchange interaction. For each disorder realization in the chemical potential, the value at each site was drawn from a Gaussian distribution centered at $\mu = 2.0t$ with standard deviation $\sigma_{\mu}$. The disorder strength was controlled by varying $\sigma_{\mu}$ up to $5\delta$, where $\delta$ is equal to half the electronic gap of the unperturbed system. Similarly, the disorder in the exchange interaction was modeled with a Gaussian distribution centered at $J = 1.92t$ and the standard deviation $\sigma_{J}$ was also varied up to $5\delta$.

\begin{figure}[t!]
\centering
\includegraphics[width=0.7\columnwidth]{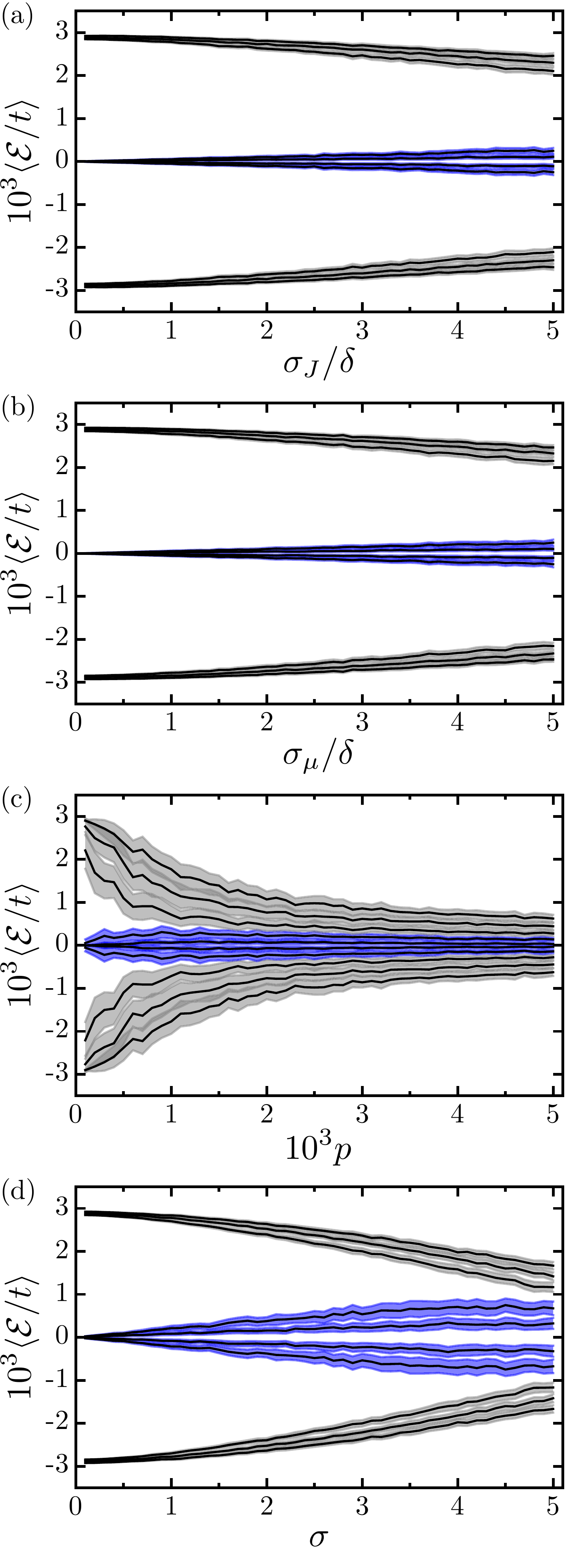}
\caption{Effect of disorder on energy states in the gap vicinity. Disorder-averaged energies (black curves) with shaded standard deviations as functions of the disorder strength parameter. Blue/Gray shaded curves correspond to in-gap/bulk states. Disorder in the chemical potential $\mu$ and exchange interaction $J$ is shown in (a) and (b), respectively. Disorder arising from deformations of the pristine magnetic texture: (c) random flips of the magnetic moments, and (d) small random distortions about the magnetic moment directions.}
\label{SpectrumDisorder}
\end{figure}

Random flips of the magnetic moments of the skyrmionic texture were modeled using the Bernoulli distribution with probability parameter $p$, which gives a value of 1 with probability $p$ or a value of 0 with probability $1 - p$. For a given disorder realization, a value drawn from the Bernoulli distribution was assigned to each magnetic moment site. Only the moments with assigned value of 1 were flipped. In this case the disorder strength was controlled by $p$, which was swept in the interval $[0.0 , 5 \times 10^{-3}]$.

Small distortions throughout the entirety of the pristine magnetic texture were constructed as follows. If $\Bn_{\Br}$ is the direction of the magnetic moment at site $\Br$, then adding a small random vector $\Bv_{\Br}$, whose components are drawn from a Gaussian distribution with zero mean and standard deviation $\sigma_{\rm{rad}}$, results in a vector whose direction is within a cone with axis $\Bn_{\Br}$ and cone angle $\approx \sigma_{\rm{rad}}$ (in radians). Therefore, at each site the new distorted direction $\Bn_{\Br}'$ is given by
\begin{align}
\Bn_{\Br}' = \frac{\Bn_{\Br} +\Bv_{\Br}}{|\Bn_{\Br} +\Bv_{\Br}|} \,.
\end{align} 
The disorder strength was in this case controlled by varying $\sigma_{\rm{rad}}$. It is useful to define $\sigma = (180/\pi) \sigma_{\rm{rad}}$ which corresponds to an angle in degrees. The parameter $\sigma$ was increased from $0^{\circ}$ up to $5^{\circ}$.

Supplementary Figure~\ref{SpectrumDisorder} shows the effect of the four types of disorder on the energy states in the vicinity of the gap. The disorder strength parameters of each disorder type are $\sigma_{\mu}$, $\sigma_{J}$, $p$, and $\sigma$. For each value of these parameters, the spectrum was computed for a total of 100 disorder realizations. The disorder-averaged energy (black curves) and standard deviation (shades) of each state were then calculated. We conclude that MBSs are robust against disorder in the electronic couplings $\mu$ and $J$, and they are more susceptible to deformations in the magnetic texture.

%%%%%%%%%%%%%%%%%%%%%%%%%%%%%%%%%%%%%%%%%%%%%%%%%%%%%%%%%%
%%%%%%%%%%%%%%%%%%%%%%%%%%%%%%%%%%%%%%%%%%%%%%%%%%%%%%%%%%

\section{Supplementary Note 3: One-dimensional spin helices}

\begin{figure}[t!]
\includegraphics[width=\columnwidth]{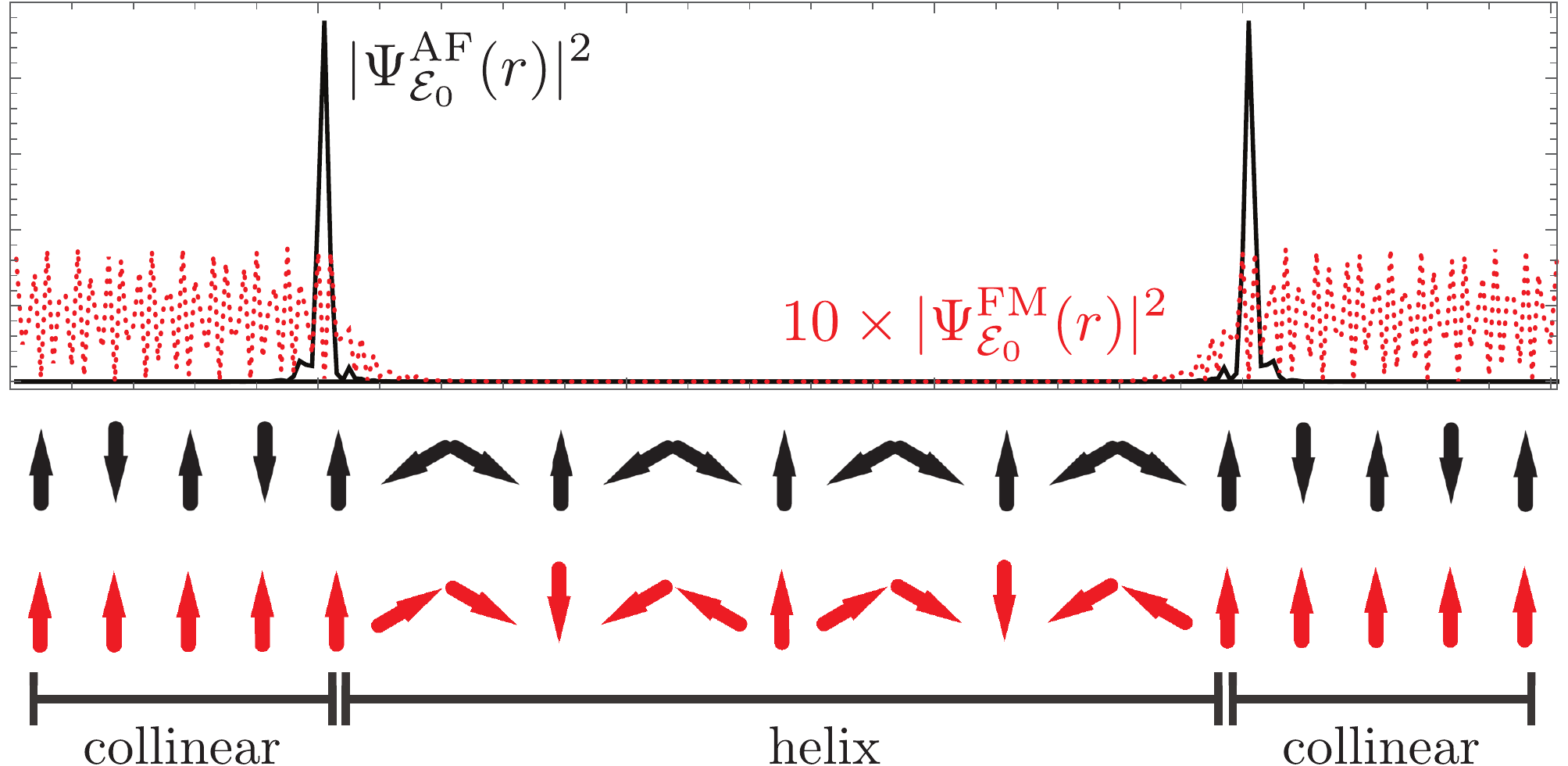}
\caption{Localization of zero-energy states in one-dimensional spin helices. The probability density of the zero-energy states, $|\Psi^\textrm{AF}_{\CalE_0}(r)|^2$ and $|\Psi^\textrm{FM}_{\CalE_0}(r)|^2$, is localized at the interface between antiferromagnetic helical and collinear antiferromagnetic order (black solid curve) but delocalized over the collinear ferromagnetic region (red dotted curve) for ferromagnetic helices. Both helices extend over twenty-five times their pitch which has a length of six lattice sites while the material parameters are $\{\mu,\Delta,J\}=\{3.0 t,0.5 t,2.5 t\}$.
}
\label{leak}
\end{figure}

The necessary conditions to host MBSs in superconductors with a noncollinear or effectively noncollinear \textit{ferro}magnetic field have been extensively discussed in the literature of one-dimensional topological materials~\cite{klinovajaPRL13,vazifehPRL13,brauneckerPRL13}. Perhaps less well known are the conditions necessary with a noncollinear \textit{antiferro}magnetic field~\cite{hsuPRB15}. Here we use our model Hamiltonian [Eq.~\eqref{ham} in the main text] to characterize a few of the important properties of one-dimensional spin helices. The spins reside at lattice sites $\Br = i \xhat$ with $i$ an integer, and the lattice constant is taken as unity for simplicity. The texture of the helix is given by $\Bn_{\Br}=(\mathcal P)^{i} [0 , \sin(k_h r) , \cos(k_h r)]^T$, where $r = |\Br|$, $\lambda_h=2\pi/k_h$ is the helix pitch, and $\mathcal P=\pm1$ for a ferromagnet or antiferromagnet, respectively. Upon transforming to momentum space, $k$, the spectrum of the infinite antiferromagnetic helix exhibits a gap closing at $k=\pi$ and supports a topological phase when $\sqrt{\Delta^2+[2t\sin(k_h /2)-|\mu|]^2}<|J|<\sqrt{\Delta^2+[2t\sin(k_h /2)+|\mu|]^2}$. This is in contrast to the ferromagnetic helix in which the gap closes at $k=0$ and is topological when $\sqrt{\Delta^2+[2t\cos(k_h /2)-|\mu|]^2}<|J|<\sqrt{\Delta^2+[2t\cos(k_h /2)+|\mu|]^2}$. In the antiferromagnetic (ferromagnetic) case, the region in parameter space in which the system is topological decreases (increases) with increasing pitch. Critically, this implies that a collinear antiferromagnet, $k_h\rightarrow0$,  is always nontopological and the spectrum exhibits a trivial gap closing when $J^2=\Delta^2+\mu^2$; in contrast, the spectrum of a collinear ferromagnet is gapless when $\sqrt{\Delta^2+(2t-|\mu|)^2}<|J|<\sqrt{\Delta^2+(2t+|\mu|)^2}$.

The consequences of this are immediately realized upon considering long helices, containing many turns, flanked by a collinear region $\Bn_{\Br}=(\mathcal P)^{i}(0 , 0 , 1)^T$. If the helical part of the system is in the topological phase, we find zero-energy states at the ends of the helical textures. In an antiferromagnet, the zero-energy states are localized to the interface between the helix and collinear textures [Supplementary Figure~\ref{leak} (black solid line)]. In a ferromagnet, the zero-energy states are totally delocalized throughout the collinear ferromagnetic region  [Supplementary Figure~\ref{leak} (red dashed line)]; paradoxically, the large exchange term that drives the helical portion into the topological phase is simultaneously responsible for destroying the gap in the collinear ferromagnetic region.

%%%%%%%%%%%%%%%%%%%%%%%%%%%%%%%%%%%%%%%%%%%%%%%%%%%%%%%%%%
%%%%%%%%%%%%%%%%%%%%%%%%%%%%%%%%%%%%%%%%%%%%%%%%%%%%%%%%%%

\section{Supplementary Note 4: Antiferromagnetic skyrmion chains vs. one-dimensional spin helices}

\begin{figure}[t!]
\includegraphics[width=\columnwidth]{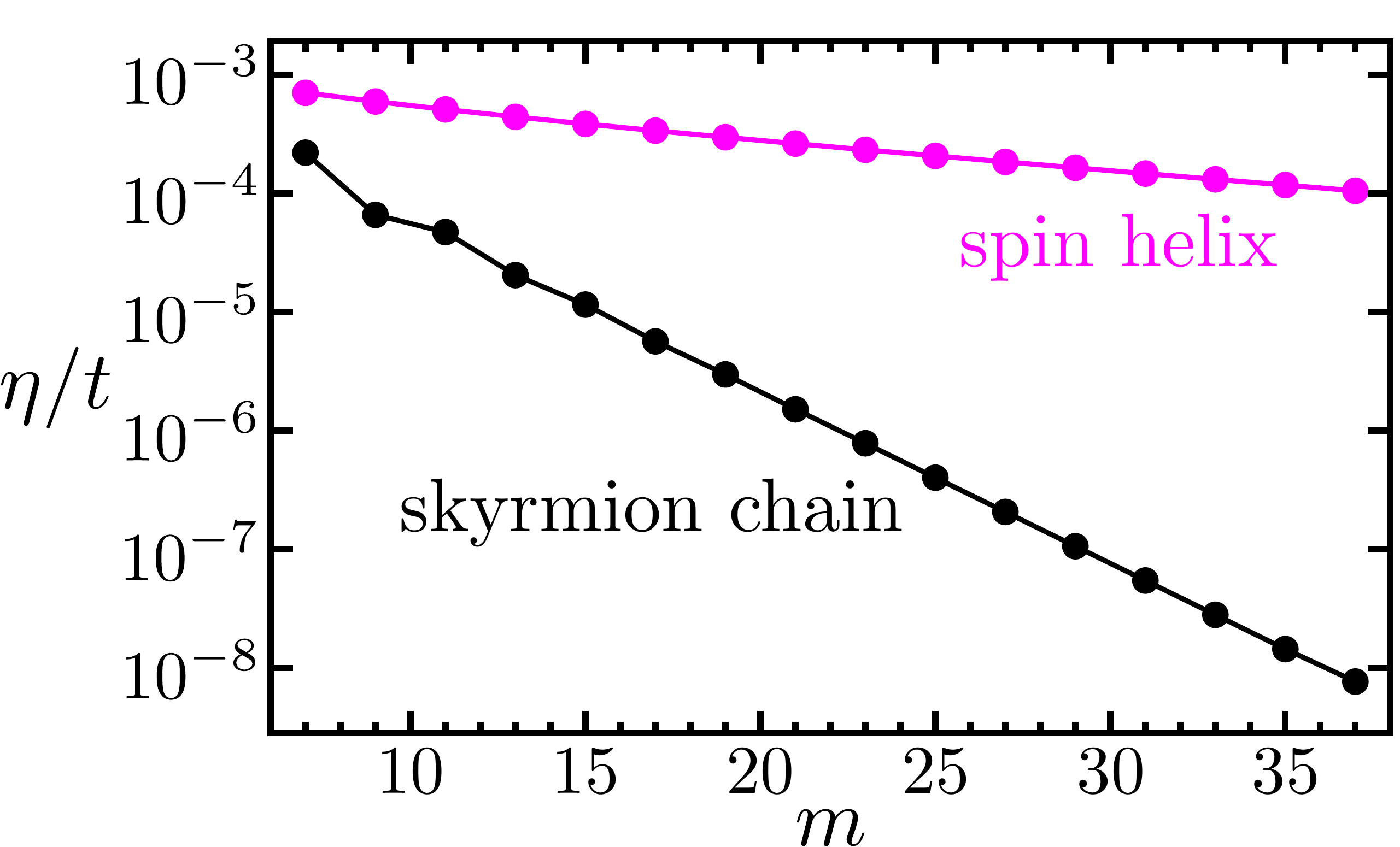}
\caption{Hybridization of MBSs. Energy difference, $\eta$, between the lowest nonnegative energy state and its particle-hole partner supported by chains comprised of $m$ antiferromagnetic skyrmions (black) and the corresponding one-dimensional antiferromagnetic spin helix (magenta) running along the chain axis. The minimum length necessary to ensure well-localized MBSs is smaller in antiferromagnetic skyrmion chains than in one-dimensional antiferromagnetic spin helices.
}
\label{comparison}
\end{figure}

There is an advantage of great importance of ASCs over one-dimensional antiferromagnetic spin helices: under equivalent conditions, the minimum length necessary to host well-localized MBSs is shorter for ASCs. We compare the realistic ASC, obtained as explained in Methods, with the line of spins running along the chain, through the center of the skyrmions, which itself is an antiferromagnetic spin helix. In order to make a meaningful comparison devoid of finite-size effects the induced superconducting gaps of the corresponding periodic textures must be the same.  (We note that we do not determine the superconducting gap self-consistently in this work~\cite{Balatsky2006,Meng2015}.) We extract a magnetic unit cell, comprised of two antiferromagnetic skyrmions, from within the bulk of the ASC to construct the corresponding electronic band structures and obtain the induced gaps. For the electronic model parameters $\{\mu,\Delta,J\}=\{2.5 t,0.5 t,2.43 t\}$ the open ASC supports two MBSs, as shown in Fig.~\ref{pd} in the main text. Fixing $\mu=2.5 t$ and $\Delta=0.5 t$, we find that for $J=1.973 t$ the induced gap of the antiferromagnetic spin helix matches that of the ASC.

Naturally, to study the localization of the MBSs we require finite textures. Starting from an open chain of 37 antiferromagnetic skyrmions we sequentially remove two skyrmions from within the bulk and compute the electronic spectra supported by the ASC and the antiferromagnetic spin helix using the parameters identified above. If the texture is not sufficiently long, the MBSs from each end may overlap and hybridize~\cite{Rainis2013}, hence acquiring a finite, nonzero energy. Therefore, as a proxy for the MBSs localization length we use the energy difference, $\eta$, between the lowest nonnegative energy state and its particle-hole partner, which we plot in Supplementary Figure~\ref{comparison}. This energy difference is larger for the antiferromagnetic spin helix than the ASC. While a chain of 13 antiferromagnetic skyrmions already supports well-localized MBSs, the antiferromagnetic spin helix of a length equivalent to 37 skyrmions exhibits considerable hybridization. The reason can be traced back to the spatial extension of the ASC perpendicular to the chain axis which allows the MBSs to spread out laterally, as shown in Fig.~\ref{pd} in the main text, and thus shorten their localization length. We conclude that for the same induced superconducting gap, the minimum length necessary to ensure well-localized MBSs is smaller in ASCs than in one-dimensional antiferromagnetic spin helices.

%%%%%%%%%%%%%%%%%%%%%%%%%%%%%%%%%%%%%%%%%%%%%%%%%%%%%%%%%%
%%%%%%%%%%%%%%%%%%%%%%%%%%%%%%%%%%%%%%%%%%%%%%%%%%%%%%%%%%

\end{document}